\documentclass[a4paper,twoside]{article}
\newcommand{\affil}[1]{$^{\rm #1}$}
\usepackage[authoryear]{natbib}
\bibpunct{(}{)}{;}{a}{}{,}
\usepackage{graphicx}
\date{Accepted for Publication Feb 2008.} 
\outer\def\gtae {$\buildrel {\lower3pt\hbox{$>$}} \over 
{\lower2pt\hbox{$\sim$}} $}

\title{\large\bf\flushleft The Coherent Radio Emission from the RS CVn 
Binary HR 1099}
\author{\parbox{\textwidth}{\flushleft
\vspace{-0.5cm}
{\it O. B. Slee\affil{A, C}, W. Wilson\affil{A}, G. Ramsay\affil{B}}\\
\vspace{0.4cm}
{\small \affil{A}\,Australia Telescope National Facility, CSIRO, P.O. 
Box 76 Epping, NSW 2121, Australia}\\
{\small \affil{B}\,Armagh Observatory, College Hill, Armagh, BT61 9DG, 
Northern Ireland, UK}\\
{\small \affil{C}\,Email: bruce.slee@csiro.au}}}
\begin{document}
\onecolumn
\begin{changemargin}{.8cm}{.5cm}
\begin{minipage}{.9\textwidth}
\vspace{-1cm}
\maketitle
%
%
\small{\bf Abstract: 

The Australia Telescope was used in March--April 2005 to observe the
1.384 and 2.368 GHz emissions from the RS CVn binary HR 1099 in two
sessions, each of 9 h duration and 11 days apart. Two intervals of
highly polarised emission, each lasting 2-3 h, were recorded. During
this coherent emission we employed a recently installed facility to
sample the data at 78 ms intervals to measure the fine temporal
structure and, in addition, all the data were used to search for fine
spectral structure. We present the following observational results;
(1) $\sim$100\% left hand circularly polarised emission was seen at
both 1.384 and 2.368 GHz during separate epochs; (2) the intervals of
highly polarised emission lasted for 2--3 h on each occasion; (3)
three 22 min integrations made at 78 ms time resolution showed that
the modulation index of the Stokes V parameter increased monotonically
as the integration time was decreased and was still increasing at our
resolution limit; (4) the extremely fine temporal structure strongly
indicates that the highly polarised emission is due to an
electron-cyclotron maser operating in the corona of one of the binary
components; (5) the first episode of what we believe is ECME
(electro-cyclotron maser emission) at 1.384 GHz contained a regular
frequency structure of bursts with FWHM $\sim$48 MHz which drifted
across the spectrum at $\sim$0.7 MHz min$^{-1}$. Our second episode of
ECME at 2.368 GHz contained wider-band frequency structure, which did
not permit us to estimate an accurate bandwidth or direction of drift;
(6) the two ECME events reported in this paper agree with six others
reported in the literature in occurring in the binary orbital phase
range 0.5 -- 0.7; (7) in one event of 8 h duration, two independent
maser sources were operating simultaneously at 1.384 and 2.368 GHz. We
discuss two kinds of maser sources that may be responsible for driving
the observed events that we believe are powered by ECME. One is based
on the widely reported `loss-cone anisotropy', the second on an
auroral analogue, which is driven by an unstable `horseshoe'
distribution of fast-electron velocities with respect to the magnetic
field direction. Generally, we favour the latter, because of its
higher growth rate and the possibility of the escape of radiation
which has been emitted at the fundamental electron cyclotron
frequency. If the auroral analogue is operating, the magnetic field in
the source cavity is $\sim$500 G at 1.384 GHz and $\sim$850 G at 2.368
GHz; the source brightness temperatures are of the order $T_{B}$
$\sim10^{15}$ K. We suggest that the ECME source may be an aurora-like
phenomenon due to the transfer of plasma from the K2 subgiant to the
G5 dwarf in a strong stellar wind, an idea that is based on VLBA maps
showing the establishment of an 8.4 GHz source near the G5 dwarf at
times of enhanced radio activity in HR 1099.  }


\medskip
\medskip
\end{minipage}
\end{changemargin}

\small

\section{Introduction}

It has been known for some years that the very active RS CVn binary HR
1099 occasionally emits highly polarised, narrow-band emission at L
(1.4 GHz) and S bands (2.4 GHz), (e.g. White and Franciosini 1995,
Jones et al. 1995).  Similar episodes have been recorded in the
emissions of some dMe flare stars, a notable example being that from
Proxima Centauri (Slee, Willes and Robinson 2003). Recently, periodic
bursts of coherent emission have been detected from an ultra-cool
dwarf star by Hallinan et al. (2007). In addition, attempts have been
made over the past few years to detect coherent radio emission from
double degenerate binary systems (Wu et al. 2002) and in degenerate
star-planet systems (Willes \& Wu 2004). In these systems, the authors
suggest that unipolar induction (UI) could be a driving mechanism,
comparable to that which drives the Jupiter-Io decametric
emission. Evidence for such emission has been detected in the
candidate double degenerate system RX J0806+15 (Ramsay et al. 2007).

Recent observations by Osten and Bastian (2006, 2007) of the highly
polarised emission from the dwarf flare star AD Leo, using time
resolutions of 10 and 1 ms respectively, showed a rich variety of
frequency and temporal structure, which ranged from diffuse bands to
narrow-band, fast-drift striae. On one occasion the authors conclude
that the mechanism is coherent plasma emission, while in another flare
at a later date they favour the electron-cyclotron maser (ECME).

By analogy with the short highly polarised solar bursts containing
structure as short as 40 milli-seconds (Dulk 1985), the stellar
emissions have been attributed to coherent sources in stellar coronae,
but up to now synthesis instruments have not possessed the millisec
time resolution to properly compare stellar and solar coherent
emissions. In the solar case, the shortest spike-like bursts are often
attributed to electron cyclotron maser emission (ECME), while the
coherent emissions responsible for the much longer bursts of Types I -
V are due to plasma emission (c.f. Dulk 1995). It is clear that in
order to distinguish between these two types of coherent emission from
stars, one needs an improvement in the time resolution of synthesis
telescopes of two orders of magnitude plus the ability to track the
bursts of emission in frequency over at least 100 MHz.

During 2004 one of us (WW) modified the correlator for the compact
array (ATCA) of the Australia Telescope to deliver samples at the rate
of 12.8 s$^{-1}$, which is 128 times faster than that used for earlier
work on stellar emissions with the ATCA. The on-line display software
(VIS) was also modified to display the flux density in the Stokes V
parameter (circularly polarised component of Stokes I); this prompts
the observer to switch to the high sampling rate of 12.8 s$^{-1}$ only
at times when highly circularly polarised emission of sufficient
intensity is seen on the display.

This observing system was utilized during two long observing sessions
with the ATCA on March 28 and April 08, 2005, and resulted in our
confirming the detection of strong coherent emission on both
occasions.

\section{Observations}

We observed HR 1099 with the ATCA in a 6 km configuration for 9 h on
both 28/3/05 and 8/4/05, recording simultaneously at 1.384 and 2.368
GHz with the usual integration time of 10 s. The flux density
calibrator was B1934-638 and the phase calibrator (interleaved with
the target integrations) was J0336-019. After the calibration, the
full data-set was mapped to check that the radio image of HR 1099 was
not contaminated by side lobes of surrounding field sources. Since HR
1099 is situated near the celestial equator, the E--W configuration of
the ATCA resulted in very poor angular resolution in declination, so
that we needed to be sure that the side lobes of surrounding field
sources were minimized. The left hand panel of Figure 1 shows a
cleaned map within a square area of 26$^{'}\times 26^{'}$ around HR
1099; the restoring beam with FWHM 569$^{''}\times 3.8^{''}$ and
major axis in PA = 0 deg is depicted in the extreme lower left. It is
clear the u-v data forming this map are likely to be contaminated by
the numerous field sources visible in this map. We therefore modelled
the field using the clean components from Figure 1 (left hand panel)
and subtracted the modelled field to produce the much better map of
Figure 1 (right hand panel), which has a dynamic range of 182 (maximum
/ minimum contour levels); the measured rms level in clear areas
around HR 1099 is $\sim$70 $\mu$Jy/beam. It is especially pleasing
that the maximum side lobes from HR 1099 itself are only 0.6\% of its
peak flux density. This modelling procedure was used at 1.384 GHz and
2.368 GHz on the data for both March 28 and April 08; the resulting
modified uv data sets were used in the following analysis.

The averaged total flux density (Stokes I) of each 25 min integration
was then measured using the MIRIAD task UVFIT and plotted against the
mid-UT of each integration. A similar set of measurements was made and
plotted for the circularly polarized component Stokes V. So far we
have been referring only to the standard 10 s integrations.

Our monitoring of the on-line display had enabled us to recognize the
presence of strong, highly circularly polarised emission. During the
one or two 25 min integrations when this emission was strongest, we
switched to the alternative high-resolution system, the results being
available in a separate data file. These data were also separately
calibrated in MIRIAD, the calibrators having also been sampled at the
rate of 12.8 s$^{-1}$. To analyse this data, we devised special
software that enabled us to apply the task UVFIT to six sets of
integrations automatically, with sampling intervals that varied
between 0.078 and 40 s. This resulted in six sets of flux densities
together with their rms fitting residuals.

\begin{figure}
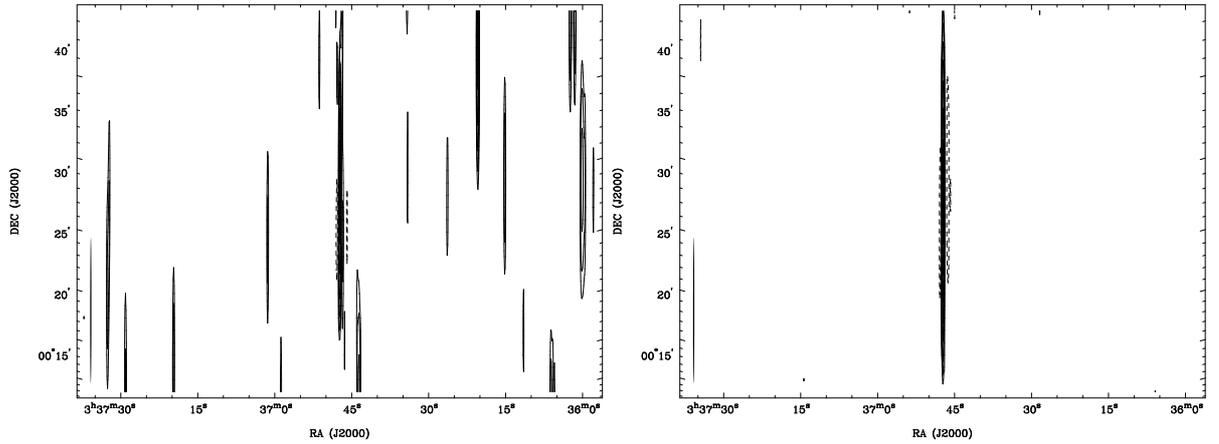

\begin{center}
\includegraphics[scale=0.3, angle=-90]{HR1099_PASA.Fig1a.ps}
\includegraphics[scale=0.3, angle=-90]{HR1099_PASA.Fig1b.ps}
\caption{Left): A cleaned total intensity map at 1.384 GHz of the
field centred on HR 1099 on March 28 2005, constructed from nineteen
25 min integrations with the data sampled at 10 s intervals. The
highest contour level is 33.62 mJy/beam and the lowest is 0.93 mJy/beam. 
The restoring beam of FWHM = 569$\times$3.8 arcsec is shown in
the extreme lower left corner. Right): The field surrounding HR 1099
after modelling the field in the left hand panel (with HR 1099 masked)
and subtracting the uv components of the field sources. The highest
contour is 33.73 mJy/beam and the lowest is 0.21 mJy/beam. The rms
over the clear area surrounding the central unresolved plot of HR 1099
is ~ 70 $\mu$Jy/beam. The restoring beam of FWHM = 569$\times$3.8
arcsec is shown in the extreme lower left corner.}
\label{fig1}
\end{center}
\end{figure}

In the present experiment we needed to examine the spectral
distribution of the continuum across the bandwidth of the receiver at
two frequencies centred on 1.384 and 2.368 GHz. These frequencies were
recorded simultaneously, with the four Stokes parameters available at
each frequency.

The IF output consisted of 13 independent frequency channels, each of
8 MHz width and spaced 8 MHz apart, permitting us to display the
spectrum over a total width of 104 MHz, resulting in total fractional
bandwidths at the lower and higher frequencies of 0.075 and 0.044
respectively. The fractional bandwidths of the thirteen 8 MHz IF
channels were 0.00578 and 0.00338 at the two frequencies.  We describe
the observations separately for each of the recording dates below.

\subsection{Observations of March 28} 

Figure 2 shows the mean flux densities of the 25 min integrations,
using the standard 10 s sampling interval.  First, the Stokes I
measurements in panel (a) at 1.384 and 2.368 GHz display markedly
different variability, with the lower frequency showing increases that
are not reproduced at 2.368 GHz. The higher frequency shows a steady
intensity of 35--37 mJy, while the 1.384 GHz emission displays an
increased level over the first 4h and a much larger increase in the
last 2 h. A glance at panel (b) suffices to show that the polarised
intensity (Stokes V) at 2.368 GHz is close to zero while the same
marked 1.384 GHz peaks that were seen in panel (a) are clearly
reproduced in panel (b). These peaks are reproduced in the fractional
circular polarisation at 1.384 GHz shown in panel (c). Figure 2
clearly demonstrates that we are observing relatively narrow-band
highly polarised emission at 1.384 GHz, but the full extent of the
fractional polarisation will not be evident until the full band width
of 104 MHz, utilized in the results of Figure 2, is subdivided into
its thirteen 8 MHz-wide frequency channels.

The intensity of the first polarised section at 1.384 GHz, shown in
panel (b) of Figure 2, was not high enough to actuate the
high-resolution mode of operation; this was brought into operation for
one of the 25 min integrations near the strong peak at 08:08 UT.

\subsection{Analysis of high time-resolution data}

Our analysis of the high time-resolution data was intended to
elucidate the temporal structure of the highly polarised emission. We
did this by allocating the data to six bins containing the samples
integrated over 40, 10, 2.5, 0.625, 0.156 and 0.078 s and then finding
the average Stokes V intensity in each bin. Call the mean flux density
associated with each bin, $<S_{t}>$, say, and its variance
$\sigma_{t}^{2}$, where t signifies the integration time of samples
assigned to that bin. In addition, the $i$th individual flux density
in that bin has its residual mean square value, $\sigma_{i}^{2}$ ,
which is dependent on the system noise and will also make a
contribution to $\sigma_{t}^{2}$. Therefore, to compute the variance
due to the star's intensity variability {\sl between} samples in the
bin, $\sigma_{t^{'}}^{2}$, we have:

\begin{equation}
\sigma_{t^{'}}^{2} = \sigma_{t}^{2}  - <\sigma_{l}^{2}>     
\end{equation}

in which the $\sigma_{l}$ have been averaged over all the samples in
the bin.

The rms value of the stellar variability is thus;   

\begin{equation}
\sigma_{t^{'}} = \{\sigma_{t}^{2} - <\sigma_{t^{'}}^{2}>\}^{1/2}
\end{equation}

Next, to assign a meaningful measure of this variability, we define a 
modulation index:                    

\begin{equation}
M=\sigma_{t^{'}}/<S_{t}>
\end{equation}

\begin{figure}
\begin{center}
\includegraphics[scale=0.6, angle=0]{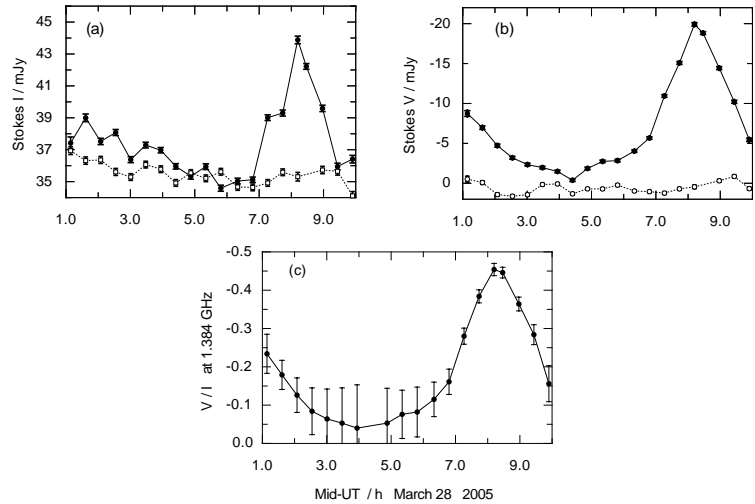}
\caption{The Stokes I and V flux densities from the 25 min integrations on  
March 28 2005 are plotted against the mid-UT of each integration. Panel (a) 
shows the I flux densities at 1.384 GHz (filled circles joined by full lines) 
and the I flux densities at 2.368 GHz (open squares joined by dotted lines).
Panel (b) shows the V flux densities at 1.384 GHz (filled circles joined by 
full lines) and at 2.368 GHz (open squares joined by dotted lines). Panel (c) 
chows the corresponding fractional values of circular polarisation (V/I) with 
their error bars. The error bars attached to the 1.384 GHz flux densities in 
panels (a) and (b) are the rms residuals from the task UVFIT; the error bars 
at 2.368 GHz are not shown but are similar in amplitude.}
\label{fig2}
\end{center}
\end{figure}

\subsection{Modulation index for March 28 2005}

The modulation indices for the coherent emission of 28 March are
plotted in Figure 3a, which shows a steadily increasing value for M as
the integration time falls to 0.078 s. It seems probable that M will
approach 100\% at still lower values of integration time, suggesting
that temporal structure as low as the 2--3 ms seen at times by Osten
and Bastian (2007) may be present. However, the presence of such fine
temporal structure does not necessarily discriminate between the
alternative mechanisms of plasma emission and ECME. Additional
attributes of the radiation such as its fractional instantaneous
bandwidth and its frequency drift rate are necessary in making this
distinction.

\begin{figure}
\begin{center}
\includegraphics[scale=0.65, angle=0]{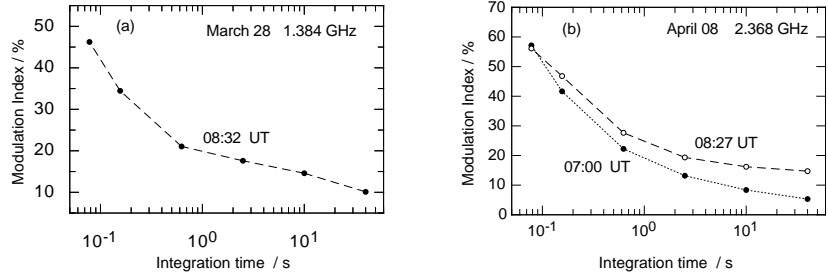}
\caption{Plots of modulation index (defined by equation (3) in Section 2.2) 
versus integration time for the high-time resolution sampling at 78 ms.
Panel (a) shows the relationship for the 1.384 GHz sample centred on 08:32 
UT on March 28   2005. Panel (b) shows the relationships derived from two 
intervals of  2.368 GHz  coherent  emission  centred on 07:00 and 08:27 UT 
on April 08  2005.}
\label{fig3}
\end{center}
\end{figure}

\subsection{Observations of April 08 2005}

Figure 4 shows the total and polarised flux densities for the 8-h
observation of April 08 in panels (a) and (b) respectively, while
panel (c) plots the polarised fraction V/I. It is evident that a
significant event occurred at 2.368 GHz in the last three hours of the
observation, but on this date there was also a slowly decreasing level
of left-handed circular polarisation at 1.384 GHz (negative Stokes
V). In this respect, the emissions differ significantly from those on
March 28, when the stonger event occurred at 1.384 GHz and a
negligible level of Stokes V was seen at 2.368 GHz.
  
The intensity levels of Stokes V at 1.384 GHz were never high enough
to actuate the high-time resolution sampling, but we were able to
record two high-resolution sections at 2.368 GHz centred on 07:00 UT
and 08:27 UT.  The Stokes I and V flux densities were similar to those
attained on March 28. The two high-resolution sections were binned in
the same manner as that described in Section 2.2 and their modulation
indices (M) were computed.

\begin{figure}
\begin{center}
\includegraphics[scale=0.65, angle=0]{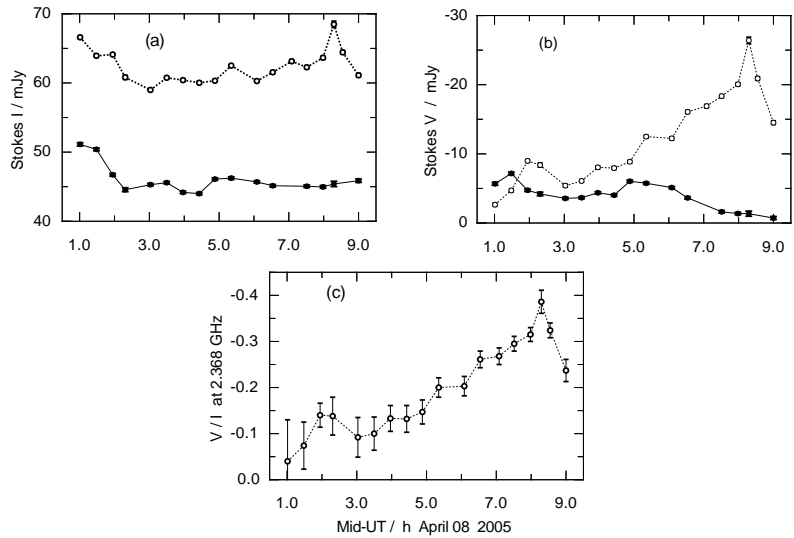}
\caption{The Stokes I and V flux densities from the 25 min integrations on 
April 08   2005 are plotted against the mid-UT  of each integration. The 
description of the three panels is identical to that of the caption to 
Figure 2.}
\label{fig4}
\end{center}
\end{figure}

\subsection{Modulation indices for April 08 2005}

Figure 3b shows that the modulation indices steadily increase with
decreasing integration time and are still increasing at our resolution
limit of 78 ms. A comparison of Figures 3a and 3b indicates that the
modulation indices of the 1.384 and 2.368 GHz coherent emissions show
similar behaviour.  Unfortunately, we do not have the capability to
increase our time and frequency resolution for continuum observations
with the existing ATNF correlator; such observations would be required
to better differentiate between possible coherent emission mechanisms.

\section{The spectral structure in coherent emission}

The IF output of the ATCA consisted of 13 independent frequency
channels, each of 8 MHz width and spaced 8 MHz apart, permitting us to
display the radio spectrum over a total width of 104 MHz. For this
task we refer to the 25 min integrations of the Stokes V data in
Figures 2b and 4b, confining our attention to the several integrations
that comprise the most highly polarised sections in these two
figures. In this spectral analysis we can utilize both low and high
time resolution integrations. The task UVSPEC in the MIRIAD software
was used to compute the averaged Stokes V and I flux densities and
their errors in each of the 8 MHz-wide channels for the selected
integrations.  These flux densities were then plotted against the
mid-frequency of each channel.

\subsection{The spectra on March 28}

Figure 5 shows the spectral structure and its dynamic behaviour at
1.384 GHz over the seven 25 min integrations that form the highly
polarised event in Figure 2b. The high time resolution data is centred
on the panel labelled 08:28 UT; here we show both the Stokes I and V
spectra. The Stokes I spectrum closely mimics the Stokes V in the
remainder of the panels. The increase in Stokes V is within a few
percent of that in Stokes I in the seven panels, with an average value
of ($\Delta{V}/\Delta{I}$) = 0.99 (weighted by variance$^{-1}$). Thus,
as far as we can determine, the emission is 100\% circularly polarised
in the left-handed sense.

The details from successive frames clearly suggest that bursts of
highly polarised emission drift relatively slowly through the total
104 MHz bandwidth at $\sim$0.7 MHz min$^{-1}$. If one fixes attention
on the burst in the top left panel, we note that by 07:16 UT it has
drifted along the frame to lower frequencies. By 07:44 UT, this burst
has practically drifted below 1.34 GHz and another burst of $\sim$45
MHz to FWHM has appeared near the high frequency end of the band. At
08:28 UT this new burst is situated near the centre of the band and by
08:58 is approaching the low frequency end. By 09:26 UT it has reached
the low frequency end of the band.

The instantaneous bandwidth of the coherent emission is one of the
critical parameters in any theoretical mechanism for its creation. The
FWHM of the most completely delineated burst at 08:28 UT is 48 MHz,
yielding a value of ($\Delta\nu/\nu$)=0.035. This, however, is an
upper limit because in the 22 min integration, from which the panel is
constructed, the centre frequency of the emission should have drifted
$\sim$15 MHz. Therefore, the true instantaneous FWHM is likely to be
$\sim$30 MHz, giving ($\Delta\nu/\nu) \sim$ 0.02. We note that Stokes
V never falls completely to zero in any of the panels, because there
are always remnants of the preceding and following bursts occupying
one or other of the ends of the 104 MHz bandwidth of the receiver. If
the channel bandwidth could be reduced to say 4 MHz, one might see
that zero Stokes V is reached between bursts, although that would
depend on how regularly they are generated.

\begin{figure}
\begin{center}
\includegraphics[scale=0.65, angle=0]{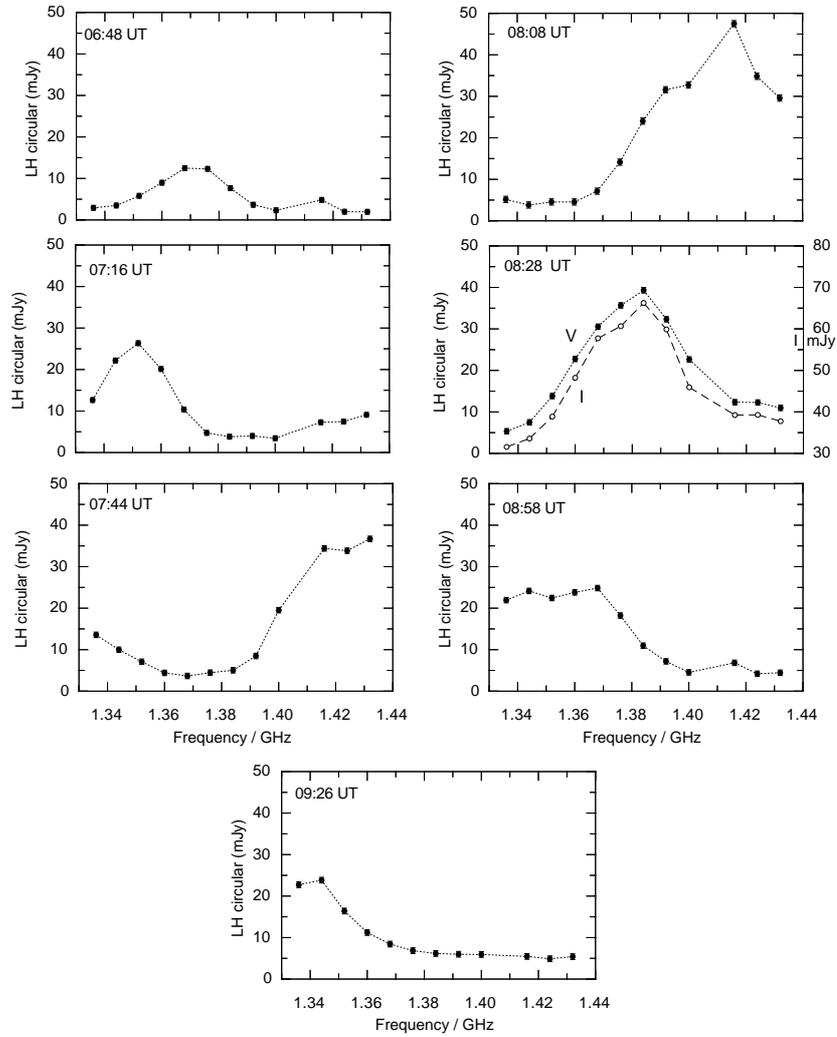}
\caption{The spectral structure in the highly polarised 1.384 GHz data
of March 28 across the 104 MHz bandwidth of the compact array. The
panels are arranged in increasing values of mid-UT of the 25 min
integrations making up the data. Six of the 7 panels show the
structure in Stokes V only, but the panel at 08:28 UT is derived from
the high-time resolution sampling and shows both the Stokes I and V
spectral structure. The frequency channel at $\sim$1.41 GHz is missing
due to an interference problem. The errors in the flux measurements are
similar to that shown in Figure \ref{fig6}.}
\label{fig5}
\end{center}
\end{figure}

\subsection{The spectra on April 08}

Figure 6 shows the spectral structure at 2.368 GHz over the nine 25
min integrations that contribute to the highly polarised event in
Figure 4b.  Data with high time resolution were recorded in the panels
labelled 07:05 UT and 08:33 UT; in the latter we show both the Stokes
I and Stokes V data on the same intensity scale spacings. The Stokes I
and V spectra are very similar in shape and amplitude on all these
scans. The larger error bars in the panel labelled 08:18 UT are due to
this integration having been cut short in order to begin the high time
resolution observations that contribute to the next panel. The
increase in Stokes V and I are very similar with an average value of
($\Delta{V}/\Delta{I}$) = 0.99 (weighted by variance$^{-1}$). Again,
as for the observations of March 28, the emission is polarised at
close to 100\% in the left-handed sense.

The successive panels of Figure 6 show that although spectral
structure is clearly present in the 2.368 GHz polarised emission, it
is difficult to assign a direction of drift. Unlike the drifting
narrow-band bursts visible at 1.384 GHz in Figure 5, no individual
burst can be completely isolated. This is no doubt due to the much
wider bandwidth occupied by the 2.386 GHz polarized bursts. The FWHM
bandwidth of these bursts is probably closer to twice our 104 MHz
bandwidth, and one would need to be able to trace the intensity
structure over at least this frequency interval to decide on its
direction of drift. One notes that the polarised emission does not
fall to near zero in any of these panels, indicating the presence of
pronounced frequency overlapping of adjacent bursts.

\begin{figure}
\begin{center}
\includegraphics[scale=0.65, angle=0]{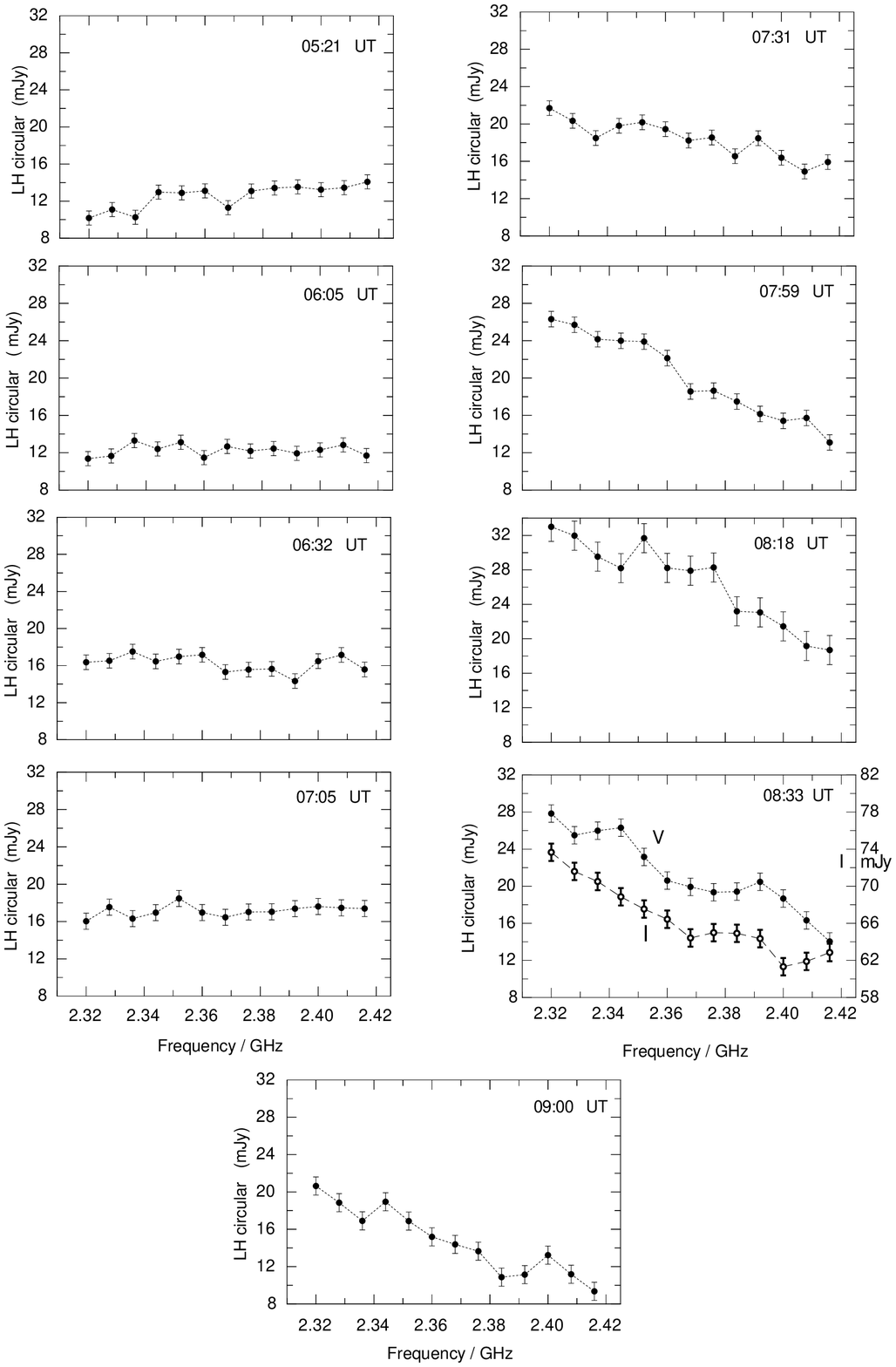}
\caption{The spectral structure in the highly polarised 2.368 GHz data of 
April 08 across the 104 MHz  bandwidth of the compact array. The panels are 
arranged in increasing values of mid-UT of the 25 min integrations making up 
the data. Eight of the nine panels show the structure in Stokes V only, but 
the panel at  08:33 UT is derived from high-time resolution sampling and 
shows the spectral variation in both Stokes I and V.  }
\label{fig6}
\end{center}
\end{figure}

\subsection{Orbital phase dependence}

VLBA observations of HR 1099 by Ransom et al. (2002) provide
reasonable evidence that during times of high radio activity the
magnetospheres of the binary components had a high degree of
interaction such that two compact 8.4 GHz sources were seen, the
stronger source centred on the K2 sub-giant and the weaker on the G5
dwarf. If our highly polarised coherent emission were to also
originate near the G5 dwarf at times of enhanced activity, one may
expect that the phase dependencies of the two phenomena would be
similar.
      
We have searched the literature to determine the epochs of other
events of long-lasting coherent emission from HR 1099 (greater than
one hour in duration and Stokes V $\sim$100\%). Table 1 gives the
result of phase binning the eight occurrences of such
emission. Although the statistics are rather poor, this table shows
that there has been a definite tendency for the coherent events to
favour the orbital phase range 0.33--0.83 with a minor peak in the
phase range 0.50--0.67. Ransom et al. (2002) used the same orbital
ephemeris to derive the phase range over which they observed the 8.4
GHz source near the G5 dwarf to move from 0.72--0.76. The secondary
source was already at full strength at the start of their
observations, so the phase range could be credibly extrapolated to
earlier phases to coincide with the peak in the coherent emission in
Table 1. This coincidence in phase range will be discussed in Section
5.

\begin{table}
\begin{center}
\begin{tabular}{lcl}
\hline
Phase & No. of Events & Ref\\
\hline
0.0--0.17 & 0 & \\
0.17--0.33 & 1 & (1)\\
0.34--0.50 & 2 & (1,2)\\
0.50--0.67 & 3 & (1,2,3)\\
0.67--0.83 & 2 & (2,3)\\
0.83--1.00 & 0 & \\
\hline
\end{tabular}
\caption{The phase dependance of the ECME events. We have used the ephemeris
of Fekel (1983) to determine the orbital phase of the events. References:
(1) Osten et al (2002), (2) Jones et al (1993), (3) this paper.}
\end{center}
\end{table}

\section{Summary of observational results}

The following observational conclusions can be drawn from our data:

\begin{enumerate}

\item Highly circularly polarised emission ($\sim100\%$ left-hand) was
seen in two observing sessions separated by 11 days.

\item The intervals of strong, highly polarised emission lasted from 2--3
hrs (Figures 2 \& 4).

\item In the first observing epoch, the highly polarised emission was seen 
only at 1.384 GHz (Figure 2).

\item During the second epoch observation, a strong, highly polarized event 
was observed at 2.368 GHz, with  weaker, highly polarised 1.384 GHz emission 
partially overlapping the stronger 2.368 GHz emission (Figure 4).

\item Three 22 min integrations were made at high time resolution (0.078 s), 
enabling us to show (Figure 3) that the modulation index of the Stokes V 
intensity increased as the integration time  was reduced and was still 
increasing at our resolution limit. 

\item During the first epoch event, coherent emission at 1.384 GHz
contained a regular frequency structure of bursts with FWHM width of
$\sim$48 MHz that drifted across the spectrum at $\sim$0.7 MHz
min$^{-1}$ (Figure 5) . The second epoch of coherent emission at 2.368
GHz also contained definite spectral structure (Figure 6), but its
significantly wider FWHM did not permit an accurate estimate of its
bandwidth nor its direction of frequency drift.

\item The two long-lasting coherent events reported in this paper conform  
with six others reported in the  literature, occurring preferentially in the 
orbital phase range of  0.50--0.67.

\end{enumerate}

\section{Discussion}

It is well known that coherent radio emission is emitted from a variety of 
celestial objects ranging from the Sun, Earth, Jupiter and Saturn to flare 
stars and some close binaries. In the case of the Sun (Melrose \& Dulk 1982), 
in the flare star AD Leo (Lang et al. 1983), and in the flare star YZ CMi 
(Lang \& Wilson 1988), high time resolution measurements have shown temporal 
structure of tens of millisec, which indicates from a light-travel time 
argument that the linear size of the emitting region should be only a few 
times $3\times10^{8}$ cm.

First, we check the peak brightness temperatures 
reached by these highly polarised bursts, using a convenient formula due to 
Osten and Bastian (2007):

\begin{equation}
T_{B} = 6 \times 10^{14} S_{mJy}(D_{pc}/\nu_{GHz} \Delta{t_{ms}})^{2}  
\end{equation}

Where $S_{mJy}$ is the peak polarised flux density, $D_{pc}$ is the
stellar distance, (29 pc) $\nu_{GHz}$ is the frequency and $\Delta t_{ms}$
is the burst duration, here taken as the time resolution of the
sampling (78 ms).

In the first epoch observation at 1.384 GHz, the maximum polarised
flux of 48 mJy (see Figure 5) results in $T_{B} > 2.1 \times 10^{15}$
K. In the second epoch observation at 2.368 GHz, the maximum polarised
flux density of $>$ 32 mJy (see Figure 6) yields $T_{B} > 4.7 \times
10^{14}$ K. If the temporal structure in these bursts is as low as
5--10 ms, the brightness temperatures would be two orders of magnitude
higher. It is clear that this completely polarised emission can not be
due to an incoherent mechanism such as that producing thermal emission
or gyro-synchrotron radiation.

Next, it is important to compare the total time intervals of coherent
emission on a M4 V flare star (AD Leo) and the close binary HR 1099 of
spectral type K3 IV + G5 V. Our Figures 2 and 3 show that the events
on HR 1099 last for 2--4 hrs as compared to ~ 1 min for the flares
reported by Osten and Bastian (2006, 2007). One questions whether this
more than two orders of magnitude difference can be compatible with
the same mechanisms operating in sources with similar linear sizes,
coronal electron densities and magnetic fields in two very dissimilar
stellar systems.

Perhaps the analysis of Osten and Bastian (2006) in their Figure 6 of
a moderate intensity, 60 s flare on AD Leo in the frequency range
1162--1568 MHz is the most suitable to compare with the results for HR
1099 in our Figure 5.  Osten and Bastian have plotted their results
with a degraded time resolution of 200 ms and a degraded frequency
resolution of 7.8 MHz. Our data are plotted with a time resolution of
78 ms and frequency resolution of 8.0 MHz.  The important conclusions
to come from this comparison are the frequency drift rates of the
bursts and their instantaneous bandwidths. Osten and Bastian find a
drift rate of 52 MHz s$^{-1}$, while for HR 1099 we measure a
frequency drift rate of $\sim$0.012 MHz s$^{-1}$, i.e. more than 3
orders of magnitude slower, although the direction of drift is from
high to low frequencies in both experiments. A second significant
difference between the two sets of data is the instantaneous bandwidth
of the bursts; in AD Leo, $\Delta\nu/\nu$\gtae 0.29, while in HR 1099,
$\Delta\nu/\nu\sim$0.036 a difference of a factor of $\sim$8. The
third outstanding difference between the above sets of data is the
difference in burst occurrence rates. In their Arecibo data of June
2003, Osten and Bastian (2006) detected only two short (60 s) bursts
in 16 hr of data, spread over four days. In our HR 1099 data of March
28 2005, we see two long-lasting intervals of coherent emission, with
the more intense episode consisting of two distinct bursts slowly
traversing our 104 MHz band width.  Our bursts may contain much finer
structure that is smoothed out by our 78 ms time resolution, but its
absence in our data makes a comparison with the higher resolution data
of Osten and Bastian (2007) a less-fruitful exercise.

Perhaps an even more relevant comparison may be made between our data
and the slowly varying decimetric coherent emission from the dMe flare
star YZ CMi (Lang \& Wilson 1988). The emission lasted for $\sim$5h
and consisted of a number of discrete bursts, each of $\sim$10 min
duration and 100\% polarised in the left-handed sense. Four of these
bursts were shown to contain narrow-band structure with a fractional
bandwidth of $\Delta\nu/\nu=$0.02, but their integration time of 10 s
did not permit the authors to investigate their temporal
structure. The authors place an upper limit of $<$ 0.05 MHz s$^{-1}$
on the frequency drift of these bursts, a value consistent with our
first epoch value of 0.012 MHz s$^{-1}$. The alternative mechanisms of
ECME and plasma emission are discussed, but no definite preference was
given.

Comparing the properties of the coherent emission which we have
detected in HR 1099 with that from other stellar systems, we favour
ECME as a more likely mechanism for our HR 1099 emission because: (1)
its 100\% polarisation is much higher than that achieved in most solar
burst of Types I--V; (2) its duration is much longer than all solar
bursts except Type IV; (3) its frequency-drift rate is orders of
magnitude lower than solar bursts and the short flares from AD Leo
reported by Osten and Bastian (2006, 2007); (4) its temporal structure
is much finer than that found in solar bursts, except perhaps that in
Type 1 solar noise storms and in the solar decimetric `spike'
emission.

Regarding the location of the coherent source in the close binary, we
must consider at least two possibilities: (1) in the corona of either
of the components; (2) a coherent source resulting from the
interaction between the magnetospheres of the components. We shall
consider both options below.

\subsection{The single-star source of coherent emission}

It is tempting to interpret our highly polarised and highly
time-structured microwave emission in terms of solar analogues, which
have been studied most extensively at metric and decimetric
wavelengths and are described comprehensively in the book `Solar
Radiophysics' edited by McLean and Labrum (1985). Whether one can
apply the models developed for solar bursts of Types I--V and their
associated continuum emissions to a binary consisting of stars of
differing spectral types is open to serious question. The solar Type I
noise storm has the long duration, high polarisation and fine
intensity structure that most resembles our emissions from HR 1099,
but this is confined to metre waves -- we do not, therefore consider
this a likely cause for the events we have detected in HR 1099.

The so-called solar `spike bursts', emitted in the frequency range
0.3--3 GHz, possess similar fine time and frequency structure to that
contained in the coherent emission from HR1099 and are highly
polarised, suggesting that such a source in the corona of either of
the binary components is a distinct possibility. The most frequently
discussed driver of solar spike bursts is an anisotropy in the pitch
angle distribution about magnetic field lines, commonly called `the
loss-cone', producing an electron-cyclotron maser.

First, it must be noted that one necessary condition for an electron
cyclotron maser to operate is that the source-region plasma has a
relatively low plasma density and/or a relatively high magnetic field
strength, such that the ratio of the electron plasma frequency,
$\nu_{p} = 8.98\times 10^{3} n_{e}^{1/2}$ MHz to the
electron-cyclotron frequency, $n_c$ = 2.80 H MHz, is small; here, H is
the magnetic field strength in gauss and $n_{e}$ the electron density
cm$^{-3}$.  For $\nu_{p}/\nu_{c} <<$ 1, the highest maser growth rate
is in the x-mode at the fundamental ($\nu\sim\nu_{p}$), but it is
highly likely that fundamental maser emission would be reabsorbed as
it propagates into harmonic absorption bands in regions with lower
magnetic field strengths. Consequently, the strongest maser emission
likely to reach the observer is x-mode emission at the second harmonic
(Melrose \& Dulk 1982). However, without knowledge of the predominant
magnetic field polarity of the active region producing the radiation,
the polarisation mode (x-mode or o-mode) cannot be ascertained from
the observed handedness. Second-harmonic emission at 1.4 GHz
corresponds to a source region field of 250 G, and the condition
$\nu_{p}/\nu_{c}<$ 1 then requires source region plasma densities $< 6
\times 10^{9}$ cm$^{-3}$. This plasma density requirement can be met
almost anywhere in the stellar corona -- even at the low coronal
height of 1.003 $R_{\odot}$ the Baumbach-Allen model of the solar
corona (Allen 1973) yields a plasma frequency of only 183 MHz, well
below either of our observing frequencies of 1.384 and 2.368 GHz.
Plasma densities will, of course, be enhanced in the converging
magnetic field lines of coronal loops near their footprints.

At this point, we may consider whether the source models developed to
account for the observed auroral kilometric radiation (AKR) could be
of assistance in interpreting the coherent sources in HR 1099. The
instructive reviews by Ergun et al. (2000), and Treumann (2006) show
that AKR electron-cyclotron emission was measured by satellites that
actually traversed the auroral source cavities, allowing detailed
theoretical models to be checked by measurements of AKR frequency,
polarisation mode, plasma frequencies, magnetic and electric field
strengths and the temporal and frequency structure of AKR.

These authors find that, although a loss cone is generated, it is
insufficient to amplify the AKR waves to levels that could account for
the high-brightness, fine structured sources within the cavity. They
propose an alternative driver, consisting of an unstable `horse-shoe'
or `shell' distribution of fast electron velocities with respect to
converging field lines in the presence of a parallel electric field
and a generally low plasma density, referring to this source region as
the `auroral cavity'. In this case, the direction of the earth's field
is known with respect to the satellite's path through the region and
so the polarisation mode can be identified. The horse-shoe
distribution is apparently a much more efficient way of increasing the
growth rate of AKR, which is preferentially emitted perpendicular to
the magnetic field lines in the right-hand circularly polarized
x-mode. Just as importantly, the fundamental of the electron-cyclotron
frequency can escape the cavity, yielding a much stronger observed
intensity of AKR than would have been observed with the loss-cone
model.

Other characteristics of the AKR that mimic the coherent emission from
HR 1099 include its fine temporal and frequency structure. Bursts of
AKR with fractional bandwidth as low as 0.01 and temporal structure of
100 ms can drift through the spectrum at various rates, but a good
example is shown in Treumann's Figure 6 . This illustrates a
consistent drift from high to lower frequencies of 7--8 kHz s$^{-1}$,
comparing well with the drift rate of the coherent event in our Figure
5 of $\sim$12 kHz s$^{-1}$ from high to lower frequencies. However, we
should be cautious in applying the auroral model to stellar coronae,
where we are dealing with magnetic fields and ambient plasma densities
that are orders of magnitude higher than auroral values.
Nevertheless, the aforementioned similarities suggest that this model
should be explored in detail for stellar coherent emission. It is
feasible that source regions having aurora-like general structure
could be found in the converging magnetic field lines of coronal
loops, but detailed modelling using realistic field strengths and
plasma densities is required to validate this proposal.

\subsection{The two-star coherent source}

If the observed coherent radiation from HR 1099 is due to the
interaction between the magnetospheres of its components it would seem
that the more active K2 subgiant could possess a strong stellar wind
capable of transferring a relatively dense plasma with imbedded
magnetic field to the corona of the G5 dwarf. Ransom et al's (2002)
VLBA maps show that a relatively weaker 8.4 GHz gyro-synchrotron
source was detected near the position of the G5 star only when the
system was in a state of high activity.  A mapping of four equal
subdivisions of the data resulted in this source moving its position
in a manner consistent with the orbital motion of the G5 dwarf about
the K2 subgiant. We have noted in section 3.3 that the orbital phases
at which the published coherent events were detected appear to favour
an orbital phase range falling within that deduced from the positions
of the secondary 8.4 GHz source.

The fact that no 8.4 GHz coherent emission was detected from either
star during this observation by Ransom et al. does not rule out its
presence at considerably lower frequencies. All occurrences of
coherent radiation from HR 1099 have been detected at frequencies
below 5 GHz, usually at about 1.4--2.4 GHz, where most lower frequency
observations have been made. Gyro-synchrotron emission of comparable
strength often coexists with the coherent emission as our Figures 2
and 4 demonstrate, but, although this does not mean that the two
radiation types necessarily emanate from the same source region in the
corona, they could have a common exciting agent.

Considering the geometry and phase dependence of the coherent events
as compared with that of Ransom et al's secondary source, we propose
that an auroral analogue may operate, in which comparatively dense
plasma is transferred from the stellar wind of the K2 subgiant into
the poloidal field of its companion. The dwarf's magnetic poles would
likely be visible, due to the low inclination (38 degrees) of the
orbital plane to the sky. Field strengths near these poles would be
considerably higher than for the Sun, due to the much more rapid
rotation rate. However, we stress again that a thorough modelling of
the situation needs to be done before the auroral analogue can be
established.

\subsection{Slow changes in the intensity and spectral distribution}

Figures 2 and 4 demonstrate that both intensity and spectral
parameters vary over an interval of nine hours and from day to day. In
March 2005 (Figure 2) the coherent event was confined to $\sim$1.4 GHz
and varied in intensity by a factor of ten between the two peaks
separated by at least 7 h. Our poor angular resolution does not permit
us to decide whether the two peaks are due to separate coronal sources
or whether the plasma density and/or magnetic field strength varies
within the one maser source. A VLBI network with an angular resolution
of $\sim$one milliarcsec at 1.4 GHz will eventually be able to decide
between these alternative interpretations.

Figure 4 (for April 2005) illustrates a phenomenon that has not been
reported hitherto. Here we observe that coherent radiation is present
simultaneously at widely separated frequencies. At $\sim$1.4 GHz its
intensity drops continuously to almost zero over the 8 h observation,
while at $\sim$2.4 GHz its intensity rises continuously to a peak near
the end of the observation; there is little, if any, correlation
between the intensities at the two frequencies, suggesting that two
separate coherent sources are operating independently and with
differing plasma densities and/or magnetic field strengths. In order
to resolve this dilemma, one is faced with the task of simultaneously
operating a VLBA network at two widely separated frequencies to
achieve an angular resolution of $\sim$one milliarcsec at the lower
frequency. Such angular resolutions could be achieved at these
decimetric wavelengths by a satellite operating with a network of
earthbound radio telescopes.

\section*{Acknowledgments} 

We thank Dr. Mark Wieringa for his modifications to the VIS software
that enabled the on-line display of both the Stokes V and I
intensities. Dr.  Vincent McIntyre created software that was essential
to the analysis of the high-time resolution data. Prof. E. Budding's
useful comments are appreciated, as were those of Dr. J. Caswell. The
insightful comments of the referee resulted in considerable
improvements to the content of the paper. The Australia Telescope
Compact Array is part of the Australia Telescope, which is funded by
the Commonwealth of Australia for operation as a National facility
managed by CSIRO.


\section{References}

Allen, C..W. 1973,  {\sl Astrophysical Quantities}, 3rd edition, 
p176, The Athlone Press\\
Dulk, G.A., 1985, Ann. Rev. Astron. Astrophys., 23, 169\\
Ergun, R.E.,  et al. 2000,  ApJ, 538, 456\\
Hallinan, G., et al. 2007, ApJ, 663, L25\\
Jones, K.L., Brown, A., Stewart, R.T., Slee, O.B. 1996, MNRAS, 283, 
1331\\
Fekel, F.C., 1983, ApJ, 268, 274\\
Lang, K.R, Wilson, R.F. 1988, ApJ, 326, 300\newline
Lang, K.R. et al. 1983, ApJ (Letters), 272, L15\newline
Osten, R.A. et al. 2004, ApJS, 153, 317\newline
Osten, R.A., Bastian, T.S. 2006, ApJ, 1016\newline
Osten, R.A., Bastian, T.S.. 2007, astro-phys, arXiv:0710.5881\newline
Melrose, D.B., Dulk, G.A. 1982, Ap.J, 259, 844\newline
McLean, D.J., Labrum, N.R. eds, 1985,  Solar Radiophysics,  CUP\newline
Ramsay, G. et al. 2007, MNRAS, 382, 461\newline
Ransom, R.R. et al. 2002, ApJ, 572, 487\newline
Slee, O.B., Willes, A.J., Robinson, R.D. 2003, PASA, 20, 257\newline
Treumann, R.A. 2006, Astron Astrophys. Rev., 13, 229\newline
White, S.M., Franciosini, E. 1995, ApJ,  444,  342\newline
Willes, A.J., Wu, K. 2004,  MNRAS, 348, 285\newline
Wu, K., Cropper, M., Ramsay, G., Sekiguchi, K. 2002, MNRAS, 331, 221\newline



\end{document}